\pdfoutput=1

\documentclass[aps,pre,preprint,groupedaddress,showpacs]{revtex4}

\usepackage{amsmath}
\usepackage{amssymb}
\usepackage{graphicx}

\begin{document}

\title{Phase-field-crystal model for liquid crystals}

\author{Hartmut L\"owen}
\affiliation{Institut f\"ur Theoretische Physik II, Weiche Materie,
Heinrich-Heine-Universit\"at D\"usseldorf, 
D-40225 D\"usseldorf, Germany}
\date{\today}

\begin{abstract} Based on static and dynamical density functional theory,  a phase-field-crystal
model is derived which involves both the translational density and the orientational degree of
ordering as well as a local director field.  The model exhibits stable
isotropic, nematic, smectic A, columnar, plastic
crystalline and orientationally ordered crystalline phases. As far as the dynamics is concerned,
the translational density is a conserved order parameter while the orientational ordering is non-conserved.
The derived phase-field-crystal model  can 
  serve for efficient numerical investigations of various nonequilibrium
situations  in liquid crystals.
\end{abstract}
%

\maketitle
%
%
\section{Introduction}
\label{sec:introduction}

Within the phase-field-crystal (PFC) model \cite{Mikko,Elder_0,Emmerich}  the crystalline density field 
is described basically in terms of a single Fourier mode, i.e. as a sinusoidal density wave.
The PFC model can be understood as a modified Landau expansion of the full 
inhomogeneous one-particle density field of the solid. 
It has been applied for large-scale numerical investigations of dynamics in the solid state. 
Characteristic examples include  calculations of a variety of different quantities:
the structure and free energy of the fluid crystal interface \cite{Axel,Tapio}, 
 crystal growth dynamics into a supercooled liquid \cite{Laszlo}, the structure \cite{Plapp}
and dynamics \cite{Voorhees} of grain boundaries, and  the Asaro-Tiller-Grinfeld instability
\cite{Asaro,Grinfeld,Elder,Voorhees2}. A solid particle just enters as a ``blob'', i.e.\ a weak density 
modulation, and the dynamics is diffusive on long time-scales, i.e.\ the density itself
is a conserved order parameter.

Recently, the PFC model was derived from dynamical density functional theory (DDFT) 
 \cite{Evans,Singh:91,Loewen:94,Sven_09}.
Static density functional theory  provides
a microscopic framework to describe crystallization in equilibrium \cite{Ramakrishnan:79,FMT1,White_bear,FMT2}
and a Landau expansion in terms of density modulations \cite{Loewen1,Ohnesorge,Lutsko,Provatas}
can be used to derive the corresponding approximative free energy for the PFC. Density functional 
theory was generalized towards nonequilibrium dynamics for Brownian systems 
\cite{Marconi,Archer_Evans,Pep} and the resulting dynamical density functional 
theory can be used to derive the dynamics of the PFC \cite{Sven_09}. First of all, this derivation 
should apply to colloidal dispersions whose short-time dynamics is clearly diffusive. 
But also molecular systems governed by Newtonian dynamics for short-times behave diffusive 
on longer time-scales and therefore the derivation might have relevance for atomic systems as well.

The PFC model has been generalized to mixtures by including more than a single density field
\cite{Provatas}. However, it has never been applied to liquid crystals which are made by particles 
with {\it orientational degrees of freedom} \cite{footnote_1}. 
Under appropriate thermodynamic conditions, these 
particles occur in liquid-crystalline phases including nematic, smectic A, columnar, and plastic crystalline
phases \cite{Frenkel,Bolhuis}. While the DDFT approach was recently generalized 
towards orientational dynamics for Brownian rods both in three \cite{Rex_08} and two \cite{Wensink_08}
dimensions, the link towards the PFC model has not yet been elaborated for orientational degrees of freedom.

In this paper, we close this gap and propose a PFC model for liquid crystalline phases. One motivation 
here is to propose a minimal model, i.e.\ the simplest nontrivial model for dynamics of liquid crystals. We 
derive this model from dynamical density functional theory. Depending on the model parameters, the resulting
model does accomodate isotropic, nematic, smectic A, columnar, plastic-crystalline phase and an 
orientational ordered crystal. It can therefore be used to describe the statics and nonequilibrium 
dynamics in various situations where these
phases are relevant. This may stimulate further numerical investigations. The model is basically 
formulated in terms of two density fields, a translational and an orientational one, plus a local director field. 
While the 
translational density field is conserved, the orientational one and the director field are nonconserved
and relax quicker.

The paper is organized as follows: in section II, we derive the PFC model from fluid-based density functional theory
by expanding the orientational dependence of the density field up to the first
nontrivial order and performing a gradient expansion in the translational coordinate. 
Then, in section III, we discuss the parameter space for which stability
of the different phases is obtained. The dynamical equations are derived from dynamical density functional theory 
in section IV. We then discuss possible 
extensions of the model to more complicated situations and give final conclusions in  section V. 

\section{Derivation of the phase-field-crystal model for liquid crystals: statics}
\label{sec:derivation}

We start our derivation from microscopic static density functional theory for liquid crystals.
We consider $N$ particles with orientational degrees of freedom described by a set of unit 
vectors $\{ {\hat u}_i; i=1,...,N \}$ and center-of-mass positions $\{ \vec{R}_i; i=1,...,N \}$.
Though most of the considerations can be done in three-dimensional space, we restrict ourselves 
in the following to two spatial dimensions, where $\vec{R}_i \in \mathbb{R}^2$ and 
 ${\hat u}_i(\phi ) = (\cos \phi, \sin \phi )$ ($\phi \in [0,2\pi[$). The system has a 
total area $A$ and is kept at finite temperature $T$.

A pair interaction potential $V( \vec{R}_1 - \vec{R}_2,{\hat u}_1, {\hat u}_2  )$ between 
two particles 1 and 2 is assumed. We henceforth consider {\it apolar\/} particles implying 
the following symmetries
\begin{align}
V( {\vec r},{\hat u}_1, {\hat u}_2  ) = V( -{\vec r},{\hat u}_1, {\hat u}_2  )= 
V( {\vec r},-{\hat u}_1, {\hat u}_2  )= V( {\vec r},{\hat u}_1, -{\hat u}_2  )= 
V( {\vec r},{\hat u}_2, {\hat u}_1  )
\end{align}
Examples for $V( \vec{R}_1 - \vec{R}_2,{\hat u}_1, {\hat u}_2  )$ comprise: i) excluded volume interactions 
as dictated by hard spherocylinders \cite{Bolhuis,Lowen_HS} or hard ellipsoids \cite{Mulder} which are
used for sterically-stabilized colloids,
ii) Yukawa segment models \cite{Yukawa_Lowen_1,Yukawa_Lowen_2,Yukawa_Lowen_3} 
used for charged colloidal rods. iii) Gay-Berne potentials \cite{Gay_Berne_1,Gay_Berne_2,Gay_Berne_3}
used for molecular liquid crystals.

The inhomogeneous one-particle density $\rho ( {\vec R}, {\hat u})$  provides the 
joint probability density to find particles at
center-of-mass-position ${\vec R}$ with orientation $\hat u$.
\begin{align}
\rho ( {\vec R}, {\hat u}) = \left\langle \sum_{i=1}^N \delta ( {\vec R} - {\vec R}_i )
\delta ( \phi - \phi_i ) \right\rangle
\end{align}
where for an observable $\mathcal{A}$
\begin{align}
\langle\mathcal{A}\rangle = {1 \over Z} \int_A d^2 R_1 ... \int_A d^2 R_N \int_0^{2\pi} d\phi_1 ...  \int_0^{2\pi} d\phi_N\; \mathcal{A}
\exp \left[ -\sum_{i,j=1; i\not=j}^N \frac{V( {\vec R}_i - {\vec R}_j,{\hat u}_i, {\hat u}_j  )}{k_BT}\right]
\end{align}
is the normalized canonical average, $k_B$ denoting Boltzmann's constant, and the classical 
canonical partition function $Z$ ensures the normalization $<1>=1$. Clearly, for apolar particles, 
$\rho ( {\vec R}, {\hat u}) = \rho ( {\vec R}, -{\hat u})$.

Classical density functional theory of inhomogeneous fluids now provides the existence of an 
excess free energy density functional such that the functional
\begin{align}
\Omega ( T, A, \mu, [\rho ( {\vec R}, {\hat u})]) = {\cal F}_{id} ( T, A, [\rho ( {\vec R}, {\hat u})])
+ {\cal F}_{exc} ( T, A, [\rho ( {\vec R}, {\hat u})]) - \int_A d^2 R \int_0^{2\pi} \mathrm{d}\phi\; \mu \rho ( {\vec R}, {\hat u})
\end{align}
is minimal for the equilibrium density field for a given chemical potential $\mu$, temperature $T$ and area $A$.
The ideal rotator gas functional ${\cal F}_{id}$ is known exactly:
\begin{align}
 {\cal F}_{id} ( T, A, [\rho ( {\vec R}, {\hat u})]) = k_B T \int_A d^2 R \int_0^{2\pi} d\phi\; \rho ( {\vec R}, {\hat u}(\phi ))
[\ln (\Lambda^2 \rho ( {\vec R}, {\hat u}(\phi ))) -1 ],
\end{align}
where $\Lambda$ denotes the (irrelevant) 
thermal wavelength. The excess free energy functional ${\cal F}_{exc} ( T, A, [\rho ( {\vec R}, {\hat u})])$, on 
the other hand, incorporates all correlations and is not known in general.
In the low density limit, a second virial approximation (Onsager functional) is getting asymptotically
exact  \cite{Frenkel}
\begin{align}
{\cal F}_{exc} ( T, A, [\rho ( {\vec R}, {\hat u})]) &\approx \frac{1}{2}\int_A d^2 R_1 \int_A d^2 R_2 
\int_0^{2\pi} d\phi_1\int_0^{2\pi} d\phi_2\notag\\
&\qquad \times\left(\exp \left(- \frac{V( {\vec R}_1 - {\vec R}_2,{\hat u}_1, {\hat u}_2)}{k_BT}\right) - 1 \right)
\rho ( {\vec R}_1, {\hat u}_1)\rho ( {\vec R}_2, {\hat u}_2).
\end{align}
More generally, the Ramakrishnan-Yussouff theory of 
freezing \cite{Ramakrishnan:79} can be applied to get the following 
 perturbative approximation for ${\cal F}_{exc}$
\begin{align}
{\cal F}_{exc} ( T, A, [\rho ( {\vec R}, {\hat u})]) &\approx -\frac{k_BT}{2} \int_A d^2 R_1 \int_A d^2 R_2 
\int_0^{2\pi} d\phi_1\int_0^{2\pi} d\phi_2\notag\\
&\qquad \times c^{(2)} ({\vec R}_1 - {\vec R}_2,{\hat u}_1, {\hat u}_2)
(\rho ( {\vec R}_1, {\hat u}_1) - {\bar \rho})(\rho ( {\vec R}_2, {\hat u}_2) - {\bar \rho})
\end{align}
which can be viewed as a truncated density expansion in the density 
difference $\rho ( {\vec R}_1, {\hat u}_1) - {\bar \rho}$ around a mean density $\bar \rho$
with the kernel representing the direct correlation function of the reference fluid
at temperature $T$ and density $\bar \rho$.

Another expression which works complementary at high density for very soft interactions \cite{Rex_08}
is a mean-field approximation
\begin{align}
{\cal F}_{exc} ( T, A, [\rho ( {\vec R}, {\hat u})]) \approx \frac{1}{2}\int_A d^2 R_1 \int_A d^2 R_2 
\int_0^{2\pi} d\phi_1\int_0^{2\pi} d\phi_2 \;
V( {\vec R}_1 - {\vec R}_2,{\hat u}_1, {\hat u}_2)
\rho ( {\vec R}_1, {\hat u}_1)\rho ( {\vec R}_2, {\hat u}_2).
\end{align}
More accurate forms for ${\cal F}_{exc}$ have been proposed for hard particles using 
weighted-density-approximations \cite{Holyst,Graf} or fundamental-measure theory \cite{FMT2}.

In the following we shall adopt the Ramakrishnan-Yussouff theory and approximate
 further by only considering weak anisotropies in the orientations. The leading 
expression in the density parametrization is then
\begin{align}\label{eq:XXX}
\rho ( {\vec R}, {\hat u}) = {\bar \rho} + {\bar \rho} \psi_1({\vec r}) 
+  {\bar \rho} \psi_2({\vec r}) \left(({\hat u}\cdot {\hat u}_0({\vec r}))^2 - {1\over 2}\right) +...
\end{align}
Here, the real-valued dimensionless orientationally averaged density is  $\psi_1({\vec r})$ which is identical
to the original treatment of the PFC model  \cite{Mikko,Elder_0}. The dimensionless field $\psi_2({\vec r})$,
on the other hand, measures the local degree of orientational order. For apolar particles, 
the leading anisotropic contribution is the third term on the right-hand-side of Equ. \eqref{eq:XXX}.
Finally, the field ${\hat u}_0({\vec r})$ defines the local director of the orientational field. \cite{footnote_Referee_B}

We now derive the static free energy functional. With
\begin{align}\label{X0}
 x = \psi_1 + \psi_2P_2(\hat{u}\cdot \hat{u}_0)
\end{align}
where $P_2(y) = y^2 - \frac{1}{2}$, the ideal rotator gas part reads as
\begin{align}
\mathcal{F}_{id} &= k_BT \bar{\rho} \int_A\mathrm{d}^2R \int_0^{2\pi}\mathrm{d}\phi\;(1+x)\left[\ln \left(\Lambda^2\bar{\rho}(1+x)\right) - 1\right]\notag\\
&= F_0 + k_BT \bar{\rho} \int_A\mathrm{d}^2R \int_0^{2\pi}\mathrm{d}\phi\left( \frac{1}{2}x^2 - \frac{1}{6}x^3 + \frac{1}{12}x^4 + \mathcal{O}(x^5)\right)\label{X1}
\end{align}
where  $F_0 = 2\pi Ak_BT\bar{\rho}(\ln (\Lambda^2\bar{\rho})-1)$ and irrelevant terms linear in $x$ on the right hand side of Equ. \eqref{X1} were absorbed in a scaled chemical potential.
Inserting \eqref{X0} and performing the angular average, we obtain
\begin{align}
	\mathcal{F}_{id}\left[\psi_1, \psi_2, \hat{u}_0\right] = F_0 + \bar{\rho}k_BT \pi\int_A\mathrm{d}^2R \; \left\{ \psi_1^2 + \frac{\psi_2^2}{8} - \frac{\psi_1^3}{3} - \frac{\psi_1\psi_2^2}{8} + \frac{\psi_1^4}{6} + \frac{\psi_1^2\psi_2^2}{8} + \frac{\psi_2^4}{256}\right\}\label{X2}.
\end{align}
The correlational part within the Ramakrishnan--Yussouff approximation is
\begin{align}
	\mathcal{F}_{exc} &= -\frac{k_BT \bar{\rho}^2}{2}\int_A\mathrm{d}^2R_1\int_A\mathrm{d}^2R_2\int_0^{2\pi}\mathrm{d}\phi_1\int_0^{2\pi}\mathrm{d}\phi_2\left(\psi_1(\vec{R}_1) + \psi_2(\vec{R}_1)P_2\left(\hat{u}(\phi_1)\cdot\hat{u}_0(\vec{R}_1)\right)\right)\notag\\
&\quad\cdot\left(\psi_1(\vec{R}_2) + \psi_2(\vec{R}_2)P_2\left(\hat{u}(\phi_2)\cdot\hat{u}_0(\vec{R}_2)\right)\right)c^{(2)}(\vec{R}_1 - \vec{R}_2,\phi_1,\phi_2)
\end{align}
We now decompose
\begin{align}
	c(\vec{R},\phi_1,\phi_2) = \sum_{m=-\infty}^\infty\sum_{m'=-\infty}^\infty c_{mm'}(\vec{R})e^{2im\phi_1}e^{2im'\phi_2}
\end{align}
and consider only the leading terms where $m,m'\in\left\{-1,0,1\right\}$. The relevant expansion coefficients are
\begin{align}
	c_{mm'}(\vec{R}) = \frac{1}{(2\pi)^2}\int_0^{2\pi}\mathrm{d}\phi\int_0^{2\pi}\mathrm{d}\phi' \; e^{-2im\phi}e^{-2im'\phi'}c^{(2)}(\vec{R},\phi, \phi').
\end{align}
By symmetry, it can be shown that $c_{00}(\vec{R})$, $c_{-11}(\vec{R})$ and $c_{1-1}(\vec{R})$ only depend on $|\vec{R}|$.
Therefore
\begin{align}
\mathcal{F}_{exc} &= -\frac{k_BT\bar{\rho}^2}{2}\int_A\mathrm{d}^2R_1 \int_A\mathrm{d}^2R_2\;4\pi^2 \left[c_{00}(|\vec{R}_1-\vec{R}_2|)\psi_1(\vec{R}_1)\psi_1(\vec{R}_2)\right.\notag\\
&\quad + \frac{1}{4}\psi_1(\vec{R}_1)\psi_2(\vec{R}_2)\left\{c_{0-1}(\vec{R}_1-\vec{R}_2)e^{-2i\phi_0(\vec{R}_2)} + c_{01}(\vec{R}_1-\vec{R}_2)e^{2i\phi_0(\vec{R}_2)}\right\}\notag\\
&\quad + \frac{1}{4}\psi_1(\vec{R}_2)\psi_2(\vec{R}_1)\left\{c_{-10}(\vec{R}_1-\vec{R}_2)e^{-2i\phi_0(\vec{R}_1)} + c_{10}(\vec{R}_1-\vec{R}_2)e^{2i\phi_0(\vec{R}_1)}\right\}\notag\\
&\quad + \frac{1}{16}\psi_2(\vec{R}_1)\psi_2(\vec{R}_2)\left\{c_{-1-1}(\vec{R}_1-\vec{R}_2)e^{-2i\phi_0(\vec{R}_1)-2i\phi_0(\vec{R}_2)} + c_{-11}(|\vec{R}_1-\vec{R}_2|)e^{-2i\phi_0(\vec{R}_1) + 2i\phi_0(\vec{R}_2)}\right.\notag\\
&\left.\left.\quad + c_{1-1}(|\vec{R}_1-\vec{R}_2|)e^{2i\phi_0(\vec{R}_1)-2i\phi_0(\vec{R}_2)} + c_{11}(\vec{R}_1-\vec{R}_2)e^{2i\phi_0(\vec{R}_1) + 2i\phi_0(\vec{R}_2)}\right\}\right].\label{X10}
\end{align}
Now a gradient expansion is performed \cite{Provatas} up to fourth order in the $\psi_1\psi_1$ term of Eqn.\ 
\eqref{X10} and up to second order in the $\psi_1\psi_2$ and $\psi_2\psi_2$ terms. We  assume that 
the highest gradient term ensures stability.
Thereby one obtains
\begin{align}\label{eq:functional}
 \mathcal{F}_{exc} = \mathcal{F}_{exc}^{(1)} + \mathcal{F}_{exc}^{(2)} + \mathcal{F}_{exc}^{(3)}
\end{align}
with
\begin{align}\label{X12}
 \frac{\mathcal{F}_{exc}^{(1)}}{k_BT} = 2\pi^2\bar{\rho}\int_A\mathrm{d}^2R\left[A\psi_1^2(\vec{R}) - B\left(\vec{\nabla}\psi_1(\vec{R})\right)^2 + C\left(\Delta\psi_1(\vec{R})\right)^2\right]
\end{align}
and
\begin{align}\label{X13}
  \frac{\mathcal{F}_{exc}^{(2)}}{k_BT} = 2\pi^2\bar{\rho}\int_A\mathrm{d}^2R\left[D\psi_2^2(\vec{R}) + E\left\{\left(\vec{\nabla}\psi_2(\vec{R})\right)^2 + 4\psi_2^2(\vec{R})\left(\vec{\nabla}\phi_0(\vec{R})\right)^2\right\}\right]
\end{align}
and
\begin{align}\label{X14}
 \frac{\mathcal{F}_{exc}^{(3)}}{k_BT} = 2\pi^2\bar{\rho}\int_A\mathrm{d}^2R\; F\left[\left(\vec{\nabla}\psi_1(\vec{R})\right)\cdot\left(\vec{\nabla}\psi_2(\vec{R})\right) + 2\psi_2(\vec{R})(\hat{u}_0(\vec{R})\cdot \vec{\nabla})^2\psi_1(\vec{R})\right].
\end{align}
In detail, in \eqref{X14}, $(\hat{u}_0(\vec{R})\cdot \vec{\nabla})^2 := \sum_{i,j=1}^2 u_{0i}(\vec{R})u_{0j}(\vec{R})\partial_i\partial_j\;$ where $\;u_{0i}(\vec{R}) = \begin{pmatrix} \cos \phi_0(\vec{R}) \\ \sin \phi_0(\vec{R}) \end{pmatrix}_{i}$.

In \eqref{X12}--\eqref{X14}, the coefficients $A,B,C,D,E$ and $F$ are generalized moments of the direct correlation function. In general,
 they depend on the thermodynamic conditions $(T,\bar{\rho})$. In detail \cite{footnote_2} (for $A=\mathbb{R}^2$),
\begin{align}
	A &= -2\pi \bar{\rho}\int_{0}^{\infty}\mathrm{d}R\; Rc_{00}(R)\\
	B &= \pi \bar{\rho}\int_{0}^{\infty}\mathrm{d}R\; R^3c_{00}(R)\\
	C &= -\frac{\pi\bar{\rho}}{12} \int_{0}^{\infty}\mathrm{d}R\; R^5c_{00}(R)\\
	D &= -\frac{\pi \bar{\rho}}{4}\int_{0}^{\infty}\mathrm{d}R\; Rc_{-11}(R)\\
	E &= \pi \bar{\rho}\int_{0}^{\infty}\mathrm{d}R\; R^3 c_{-11}(R)\\
	F &= -\frac{\pi}{8}\int_{0}^{\infty}\mathrm{d}R\; R^3 c_{01}(R\cos\phi_R,R\sin\phi_R) e^{2i\phi_R}.
\end{align}
As a remark: $F$ does not depend on $\phi_R$. For stability reasons, we henceforth assume $C,E>0$. 

Let us now discuss the static free energy functional. In the limit of no orientational order, $\psi_2 \equiv 0$, one recovers the phase--field crystal model of Elder and coworkers \cite{Mikko, Elder_0}. The expansion up to fourth order is formally similar to a Landau expansion of the smectic A--isotropic phase transition if $\psi_1$ represents the smectic order parameter \cite{Pleiner2001}. In the opposite case of constant $\psi_1$ \underline{and} constant $\psi_2$, Frank's elastic energy with a nonvanishing splay and vanishing bend modulus is recovered in the term $\sim (\vec{\nabla}\phi_0(\vec{R}))^2$ in \eqref{X13}. In fact, in two spatial dimensions there are only two Frank elastic constants since the twist modulus vanishes. If $\psi_1$ is constant and both $\psi_2$ and $\hat{u}_0$ are space dependent, we obtain the Landau--de Gennes free energy \cite{deGennes} used for inhomogeneous uniaxial nematics. In two dimensions, the Landau--de Gennes free energy has only one gradient coefficient \cite{daGama} which is  again the parameter $E$. For space--dependent $\psi_1$ and $\hat{u}_0$, but constant $\psi_2$, the free energy derived by Pleiner and Brand \cite{Brand1980} is recovered. Finally, for space--dependent $\psi_1$ and $\psi_2$, we obtain the coupling terms in $\mathcal{F}_{exc}^{(3)}$ proposed by Brand and Pleiner \cite{Brand1987}. However, the full free energy functional \eqref{eq:functional} with fourth--order gradients in $\psi_1$ and the appropriate couplings to $\psi_2$ and $\hat{u}_0$ is new and constitutes the basic static result of this paper.

\section{Equilibrium bulk phase diagram}
\label{sec:diagram}

By minimizing the free energy functional, for given thermodynamic parameters $T$ and $\bar{\rho}$, the equilibrium 
phase diagram is gained. In the special case of $\psi_2=0$, the PFC phase diagram of Elder and coworkers \cite{Elder_0} is obtained. By scaling out a length scale, there are only two remaining parameters for which a fluid, a triangular phase and an (unphysical) stripe phase is stable (see Fig. 4 in \cite{Elder_0}). For $D < -\frac{\pi}{4}$, a nonzero stable value for $\psi_2$ occurs. Combined with the PFC phase diagram, the possibility of a nematic phase and an orientationally ordered crystal emerges. The stripe phase at $\psi_2 \neq 0$ becomes either a smectic A or columnar phase depending on the sign of the parameter $F$. In fact, it was already shown in
Ref.\  \cite{vanRoij} that an Onsager-functional  yields a smectic A phase.

All the possible liquid crystalline phases are summarized in Table \ref{tab:PhaseDiag} together with their characterizing values for the number density $\psi_1$, the nematic order parameter $\psi_2$ and the director field $\hat{u}_0$. For $D > -\frac{\pi}{8}$, a plastic crystal and the ordinary isotropic phase can be stable. The full numerical calculation of the equilibrium phase diagram as a function of the parameters $A$,
$B$, $C$, $E$, and $F$ is planned in a future study.

\begin{table}
	\centering
		\begin{tabular}{|l||c|c|c|}\hline
			liquid crystalline phase	&	$\psi_1$	&	$\psi_2$	&	$\hat{u}_0$\\\hline\hline
			isotropic	&	$0$	&	$0$	&	irrelevant\\\hline
			nematic	&	constant	&	$\neq 0$	&	constant\\\hline
			plastic crystalline	&	oscillatory	&	$0$	&	irrelevant\\\hline
			orientationally ordered crystalline	&	oscillatory	&	$\neq 0$	&	constant or oscillatory\\\hline
			smectic A	&	planar oscillatory	&	$\neq 0$	&	constant, oscillatory $\parallel \vec{\nabla}\psi_1$\\\hline
			columnar	&	planar oscillatory	&	$\neq 0$	&	constant, oscillatory $\perp \vec{\nabla}\psi_1$\\\hline			
		\end{tabular}
	\caption{Characteristic values for the number density $\psi_1$, the nematic order parameter $\psi_2$ and the director field $\hat{u}_0$ for six different liquid crystalline phases, namely isotropic, nematic, plastic crystalline, orientationally ordered crystalline, smectic A and columnar.}
	\label{tab:PhaseDiag}
\end{table}

\section{Derivation of the phase-field-crystal model for liquid crystals: dynamics}
\label{sec:derivation2}

\subsection{Dynamical density functional theory} 

In two spatial dimensions, the dynamical density functional theory for Brownian systems is a 
deterministic equation for the time dependent one-particle density field
$\rho ( {\vec R}, {\hat u}, t)$ \cite{Wensink_08}:
\begin{align}\label{D1}
\frac{\partial \rho (\vec{R},\hat{u},t)}{\partial t} = \vec{\nabla}^T\cdot D_T\cdot \left(\frac{\rho(\vec{R},\hat{u},t)}{k_BT} \vec{\nabla} \frac{\delta \mathcal{F}}{\delta \rho(\vec{R},\hat{u},t)} \right) + D_R \frac{\partial}{\partial \phi}\left(\frac{\rho(\vec{R},\hat{u},t)}{k_BT} \frac{\partial}{\partial \phi} \frac{\delta \mathcal{F}}{\delta \rho(\vec{R},\hat{u},t)}\right).
\end{align}
Here, ${\bf D}_T$ is the diagonal translational short-time diffusion tensor which we assume 
to be isotropic in the following, ${\bf D}_T= diag(D_T, D_T)$, and 
$D_R$ is the rotational diffusion constant. 
Furthermore, $\mathcal{F} = \mathcal{F}_{id} + \mathcal{F}_{exc}$ is the total free energy functional. 
If the density parametrization \eqref{eq:XXX} is used, this becomes a functional $\mathcal{F}[\psi_1(\vec{R}), \psi_2(\vec{R}), \phi_0(\vec{R})]$ of the three scalar fields $\psi_1(\vec{R}), \psi_2(\vec{R}), \phi_0(\vec{R})$. Now  the chain rule of functional differentiation yields:
\begin{align}\label{eq:chainrule}
	\frac{\delta \mathcal{F}}{\delta \rho(\vec{R}, \phi)} = \frac{1}{2\pi\bar{\rho}}\frac{\delta \mathcal{F}}{\delta \psi_1(\vec{R})} + \frac{4}{\pi\bar{\rho}}\frac{\delta \mathcal{F}}{\delta \psi_2(\vec{R})}P_2(\cos(\phi - \phi_0(\vec{R}))) + \frac{1}{\pi\bar{\rho}}\frac{\delta \mathcal{F}}{\delta \phi_0(\vec{R})}\frac{\sin(2(\phi - \phi_0(\vec{R})))}{\psi_2(\vec{R})}.
\end{align}
By inserting this into Eqn. \eqref{D1}, coupled equations of motion can be obtained.

\subsection{Derivation of the dynamics (PFC1 model)}

First we describe the dynamics for the most case which is called 
PFC1 model in Ref.\ \cite{Sven_09}. The PFC1 model avoids two further approximations, namely
the expansion of 
the logarithm \eqref{X1} and a constant mobility assumption. 
By inserting the chain rule \eqref{eq:chainrule} into the dynamical density functional theory \eqref{D1}, one obtains
 dynamical equations for the three scalar fields $\psi_1(\vec{R},t), \psi_2(\vec{R},t)$ and $\phi_0(\vec{R},t)$ as follows:
\begin{align}\label{eq:time}
 k_BT\bar{\rho}\dot{\psi_1} &= k_BT\bar{\rho}D_T\Delta\psi_1 + \frac{D_T}{\pi}\left(\frac{1}{2}\vec{\nabla}\left((1+\psi_1)\vec{\nabla}\frac{\delta\mathcal{F}_{exc}}{\delta\psi_1} - \frac{\delta\mathcal{F}_{exc}}{\delta\phi_0}\vec{\nabla}\phi_0 + \psi_2\vec{\nabla}\frac{\delta \mathcal{F}_{exc}}{\delta \psi_2} \right)\right)\\
k_BT\bar{\rho}\dot{\psi_2} &= k_BT\bar{\rho}\left(D_T\Delta\psi_2 - 4D_R\psi_2 - 4D_T\psi_2\left(\vec{\nabla}\phi_0\right)^2\right)\notag\\
&\qquad + \frac{D_T}{\pi}\left(\vec{\nabla}\left(4(1+\psi_1)\left(\vec{\nabla}\frac{\delta\mathcal{F}_{exc}}{\delta\psi_2} - \frac{\delta\mathcal{F}_{exc}}{\delta\phi_0}\frac{\vec{\nabla}\phi_0}{\psi_2}\right) + \frac{\psi_2}{2}\vec{\nabla}\frac{\delta\mathcal{F}_{exc}}{\delta\psi_1}\right)\right.\notag\\
&\qquad + \left((1+\psi_1)\left(-16\frac{\delta\mathcal{F}_{exc}}{\delta\psi_2}(\vec{\nabla}\phi_0)^2 - 4\vec{\nabla}\left(\frac{\delta\mathcal{F}_{exc}}{\delta\phi_0}\frac{1}{\psi_2}\right)\vec{\nabla}\phi_0\right)\right) - \frac{D_R}{\pi}\left(16(1+\psi_1)\frac{\delta\mathcal{F}_{exc}}{\delta\psi_0}\right)\\
k_BT\bar{\rho}\psi_2\dot{\phi_0} &= k_BT\bar{\rho}\left(D_T\psi_2\Delta\phi_0 + 2D_T\left(\vec{\nabla}\psi_2\cdot\vec{\nabla}\phi_0\right)\right)\notag\\
&\qquad + \frac{D_T}{\pi}\left[\left(4(1+\psi_1)\left(\vec{\nabla}\frac{\delta\mathcal{F}_{exc}}{\delta\psi_2} - \frac{\delta\mathcal{F}_{exc}}{\delta\phi_0}\frac{\vec{\nabla}\phi_0}{\psi_2}\right) + \frac{\psi_2}{2}\vec{\nabla}\frac{\delta\mathcal{F}_{exc}}{\delta\psi_1}\right)\vec{\nabla}\phi_0\right.\notag\\
&\qquad \left.+ \vec{\nabla}\left((1+\psi_1)\left(4\frac{\delta\mathcal{F}_{exc}}{\delta\psi_2}\vec{\nabla}\phi_0 + \vec{\nabla}\left(\frac{\delta\mathcal{F}_{exc}}{\delta\phi_0}\frac{1}{\psi_2}\right)\right)\right)\right] - \frac{D_R}{\pi}\left(4(1+\psi_1)\frac{\delta\mathcal{F}_{exc}}{\delta\phi_0}\frac{1}{\psi_2}\right).
\end{align}

The right-hand-side of Eqn.\ \eqref{eq:time} clearly shows that the time-derivative $\dot{\psi_1}$ is proportional 
to a divergence of a generalized current. This implies that a generalized continuity equation holds such that the 
order parameter field $\psi_1(\vec{R},t)$ is conserved. On the other hand, this is not true for the two remaining 
orientational order parameter fields $\psi_2(\vec{R},t)$ and $\phi_0(\vec{R},t)$ which are therefore non-conserved.

The functional derivatives are local and given by
\begin{align}
 \frac{1}{4\pi^2k_BT\bar{\rho}} \frac{\delta \mathcal{F}_{exc}}{\delta \psi_1} &=  A\psi_1 + B\Delta\psi_1 + C\Delta^2\psi_1\notag\\
&\qquad - \frac{F}{2}\Delta\psi_2 + F\sum_{i,j = 1}^2\partial_i\partial_j\left(\psi_2 u_{0i}u_{0j}\right),\\
\frac{1}{4\pi^2k_BT\bar{\rho}} \frac{\delta \mathcal{F}_{exc}}{\delta \psi_2} &=  D\psi_2 - E\Delta\psi_2 + 4E\psi_2\left(\vec{\nabla}\phi_0\right)^2\notag\\
&\qquad - \frac{F}{2}\Delta\psi_1 + F\sum_{i,j = 1}^{2}u_{0i}u_{0j}\partial_i\partial_j\psi_1,\\
\frac{1}{4\pi^2k_BT\bar{\rho}} \frac{\delta \mathcal{F}_{exc}}{\delta \phi_0} &= -4E\psi_2^2\Delta\phi_0 + F\psi_2\sum_{i,j = 1}^{2}\frac{\partial u_{0i}u_{0j}}{\partial \phi_0}\partial_i\partial_j\psi_1\\
&\qquad \text{where }\;\frac{\partial u_{0i}u_{0j}}{\partial \phi_0} = \begin{pmatrix} -\sin 2\phi_0 & \cos 2\phi_0 \\ \cos 2\phi_0 & \sin 2\phi_0 \end{pmatrix}_{ij}.
\end{align}

Combining these equations yields explicit deterministic and coupled equations of motion for the three 
order parameter fields $\psi_1(\vec{R},t), \psi_2(\vec{R},t)$ and $\phi_0(\vec{R},t)$ which can be implemented for a
numerical solution.

\subsection{Derivation of the phase-field-crystal model with constant mobility (PFC2 model)}

In the constant mobility approximation, the prefactor in front of the density functional derivatives 
on the right-hand-side of Eqn.\ \eqref{D1} is replaced  by
the constant $\frac{\bar{\rho}}{k_BT}$. Then the dynamical density functional equations simplify to
\begin{align}\label{D2}
\frac{\partial \rho (\vec{R},\hat{u},t)}{\partial t} = \left( D_T\Delta + D_R\frac{\partial^2}{\partial \phi^2}\right) \frac{\bar{\rho}}{k_BT}\frac{\delta \mathcal{F}}{\delta \rho(\vec{R},\hat{u},t)}.
\end{align}
In this case, the equations of motion for the three scalar fields 
 $\psi_1(\vec{R},t), \psi_2(\vec{R},t)$ and $\phi_0(\vec{R},t)$ read as
\begin{align}\label{eq:scalar1}
	k_BT \pi \bar{\rho}\dot{\psi_1} &= \frac{1}{2}D_T\Delta\frac{\delta \mathcal{F}}{\delta\psi_1}\\
	k_BT \pi \bar{\rho}\dot{\psi_2} &= D_T\left[4\Delta\frac{\delta \mathcal{F}}{\delta\psi_2} - 16(\vec{\nabla}\phi_0)^2\frac{\delta \mathcal{F}}{\delta\psi_2} - 8\left(\vec{\nabla}\left(\frac{\delta \mathcal{F}}{\delta\phi_0}\frac{1}{\psi_2}\right)\right)\cdot \vec{\nabla} \phi_0 - 4\frac{\delta \mathcal{F}}{\delta\phi_0}\frac{\Delta \phi_0}{\psi_2}\right] - 16 D_R \frac{\delta \mathcal{F}}{\delta\psi_2}
\end{align}
and finally
\begin{align}\label{eq:scalar2}
	k_BT\pi\bar{\rho}\psi_2\dot{\phi}_0 = D_T\left[8\left(\vec{\nabla}\frac{\delta \mathcal{F}}{\delta\psi_2}\right)\vec{\nabla}\phi_0 + 4\frac{\delta \mathcal{F}}{\delta\psi_2}\Delta\phi_0 + \Delta\left(\frac{\delta \mathcal{F}}{\delta\phi_0}\frac{1}{\psi_2}\right) - 4\frac{\delta \mathcal{F}}{\delta\phi_0}\frac{1}{\psi_2}\left(\vec{\nabla}\phi_0\right)^2\right] - 4 D_R \frac{1}{\psi_2}\frac{\delta \mathcal{F}}{\delta\phi_0}.
\end{align}

The ordinary phase-field crystal model is obtained by a subsequent expansion of the ideal rotator term
\eqref{X1} up to fourth order. Following Ref.\ \cite{Sven_09}, the resulting dynamics is called PFC2 model.
In this case, the density functional derivatives are again local and given by
\begin{align}
 \frac{1}{k_BT\bar{\rho}} \frac{\delta \mathcal{F}}{\delta \psi_1} &= \pi\left(2 + 2\psi_1 - \psi_1^2 - \frac{\psi_2^2}{8} + \frac{2}{3}\psi_1^3 + \frac{\psi_1\psi_2^2}{4}\right) + 4\pi^2\left(A\psi_1 + B\Delta\psi_1 + C\Delta^2\psi_1\right.\notag\\
&\qquad \left.-\frac{F}{2}\Delta\psi_2 + F\sum_{i,j = 1}^2\partial_i\partial_j\left(\psi_2 u_{0i}u_{0j}\right)\right)\\
\frac{1}{k_BT\bar{\rho}} \frac{\delta \mathcal{F}}{\delta \psi_2} &= \pi\left(\frac{\psi_2}{4} - \frac{1}{4}\psi_1\psi_2 + \frac{1}{4}\psi_1^2\psi_2 + \frac{\psi_2^3}{64} \right) + 4\pi^2 \left(D\psi_2 - E\Delta\psi_2 + 4E\psi_2\left(\vec{\nabla}\phi_0\right)^2\right.\notag\\
&\qquad \left.-\frac{F}{2}\Delta\psi_1 + F\sum_{i,j = 1}^{2}u_{0i}u_{0j}\partial_i\partial_j\psi_1\right)\\
\frac{1}{k_BT\bar{\rho}} \frac{\delta \mathcal{F}}{\delta \phi_0} &= \left(-4E\psi_2^2\Delta\phi_0 + F\psi_2\sum_{i,j = 1}^{2}\frac{\partial u_{0i}u_{0j}}{\partial \phi_0}\partial_i\partial_j\psi_1\right).
\end{align}
The advantage of these equations is that they reduce to the dynamics of the traditional phase-field-crystal model
in the pure translational case. For a rough numerical exploration, the PFC2 model should give
the same qualitative answer as the PFC1 model. For spherical particles this was shown in  Ref.\ \cite{Sven_09}.
The dynamical equations \eqref{eq:scalar1}-\eqref{eq:scalar2} represent the main result of this paper.

\section{Conclusions}
\label{sec:conclusions}

In conclusion, we derived from static and dynamical density functional theory
phase-field-crystal equations which govern the diffuse nonequilibrium dynamics
for liquid crystalline phases. The approximations involved a two-fold: first the density
functional is approximated by a truncated functional Taylor expansion similar in spirit to the
Ramakrishnan--Yussouff theory. Then a generalized gradient expansion in the
order parameters is performed which leads to a local density functional. In addition to the traditional scalar phase-field variable $\psi_1$,
a local scalar nematic order parameter $\psi_2$ and a local nematic director field $\phi_0$
was introduced and coupled to the phase-field variable $\psi_1$. If the additional 
variables are zero, the phase-field-crystal model of Elder and coworkers \cite{Mikko,Elder_0} is recovered.
If, on the other hand, $\psi_1$ is set to zero we recover the Landau--de Gennes free energy  for uniaxial nematics 
extended by Pleiner and Brand \cite{Brand1980,Brand1987}. The proposed phase-field-crystal
model for liquid crystals allows for a wealth of stable liquid crystalline phases including
isotropic, nematic, smectic A, columnar, plastic crystalline and orientationally ordered crystals.
How the stability of these phases depends in detail on the model parameters still 
needs to be explored numerically.
The new coupled phase-field-crystal equations can be used to simulate the nonequilibrium dynamics 
of liquid crystals. Possible problems are dynamics of topological defects in the nematic phase
\cite{Muthukumar} and the formation of metastable phases at a growing interface \cite{Bechhoefer}. As the dynamics in nematic states can be obtained by using other approaches like the one in Ref. \cite{Brand1987}, the present model
may be applicable in particular to smectic films and to two--dimensional
crystalline phases.

In the present paper, the derivation of the phase-field-crystal model
was performed in two spatial dimensions. Though more tedious there is no principle 
problem in doing the same analysis in three spatial dimensions with the use of 
spherical harmonics for the orientational degrees of freedom. Moreover the present derivation 
can in principle be done to higher order in the orientational degrees of freedom.
The translational degrees of freedom can be anisotropic for the dynamical mobility matrix \cite{footnote_3}.

\begin{acknowledgments}
  I thank H. Brand, C. V. Achim, S. van Teeffelen, H. Emmerich, U. Zimmermann, R. Wittkowski,
 and T. Ala-Nissila for helpful discussions. This work has been supported by the DFG through the DFG priority program SPP 1296.
\end{acknowledgments}
\bibliography{bib/journals,bib/PFC_liquid_crystals_2}

\begin{mcitethebibliography}{53}
\bibitem[foo()]{footnote_Referee_B}
As an equivalent description, a position-dependent nematic tensor could be used
  in the density parametrization instead of using the fields $\psi_2 ({\vec
  r})$ and ${\hat u}_0 ({\vec r})$. This notation has formal advantages and
  could be helpful in order to derive generalizations of the present model to
  three spatial dimensions.\relax
\mciteBstWouldAddEndPunctfalse
\mciteSetBstMidEndSepPunct{\mcitedefaultmidpunct}
{}{\mcitedefaultseppunct}\relax
\EndOfBibitem
\bibitem[foo()]{footnote_2}
These coefficients can also be fitted to the correlation function at nonzero
  wave vector resulting in effective fit parameters.\relax
\mciteBstWouldAddEndPunctfalse
\mciteSetBstMidEndSepPunct{\mcitedefaultmidpunct}
{}{\mcitedefaultseppunct}\relax
\EndOfBibitem
\bibitem[foo()]{footnote_3}
Finally the present analysis is based on a fluid perturbation theory of the
  density functional. In principle one can also expand around another phase
  with a broken symmetry which would result in more general coupling terms than
  presented here.\relax
\mciteBstWouldAddEndPunctfalse
\mciteSetBstMidEndSepPunct{\mcitedefaultmidpunct}
{}{\mcitedefaultseppunct}\relax
\EndOfBibitem
\end{mcitethebibliography}
\bibliographystyle{rsc}
\end{document}